\documentclass[aps,prl, twocolumn,superscriptaddress,nofootinbib,preprintnumbers]{revtex4-1}
\usepackage{amsmath}
\usepackage{graphicx}
\usepackage{amssymb}
\usepackage[colorlinks,citecolor=blue]{hyperref}

\newcommand{\beq}{\begin{equation}}
\newcommand{\eeq}{\end{equation}}
\newcommand{\bea}{\begin{eqnarray}}
\newcommand{\eea}{\end{eqnarray}}
\newcommand{\met}{\not{\!\!{\rm E}}_{T}}
\newcommand{\nn}{\nonumber}
 \usepackage{ulem}

\begin{document}

\preprint{UCRHEP-T580}

\title{ New Class of Two-Loop Neutrino Mass Models with Distinguishable Phenomenology}

\author{Qing-Hong Cao}
\email{qinghongcao@pku.edu.cn}
\affiliation{Department of Physics and State Key Laboratory of Nuclear Physics and Technology, Peking University, Beijing 100871, China}
\affiliation{Collaborative Innovation Center of Quantum Matter, Beijing, 100871, China}
\affiliation{Center for High Energy Physics, Peking University, Beijing 100871, China}

\author{Shao-Long Chen}
\email{chensl@mail.ccnu.edu.cn}
\affiliation{Key Laboratory of Quark and Lepton Physics (MoE) and Institute of Particle Physics, Central China Normal University, Wuhan 430079, China}
\affiliation{Center for High Energy Physics, Peking University, Beijing 100871, China}

\author{Ernest Ma}
\email{ma@phyun8.ucr.edu}
\affiliation{ Physics $\&$ Astronomy Department and Graduate Division, University of California, Riverside, California 92521, USA}

\author{Bin Yan}
\email{binyan@pku.edu.cn}
\affiliation{Department of Physics and State Key Laboratory of Nuclear Physics and Technology, Peking University, Beijing 100871, China}

\author{Dong-Ming Zhang}
\email{zhangdongming@pku.edu.cn}
\affiliation{Department of Physics and State Key Laboratory of Nuclear Physics and Technology, Peking University, Beijing 100871, China}

\begin{abstract}\
We discuss a new class of neutrino mass models generated in two loops, and explore specifically three new physics scenarios: (A) doubly charged scalar, (B) dark matter, and (C) leptoquark and diquark, which are verifiable at the 14 TeV LHC Run-II.  We point out how the different Higgs insertions will distinguish our two-loop topology with others if the new particles in the loop are in the simplest representations of the SM gauge group. 
\end{abstract}

\maketitle

\noindent{\bf Introduction.}
The minimal particle content of the standard model (SM) of quarks and 
leptons does not allow a nonzero neutrino mass at the level of a 
renormalizable Lagrangian.  However, it has long been known~\cite{Weinberg:1979sa} 
that an effective dimension-five operator exists for obtaining a nonzero 
Majorana neutrino mass, i.e.
\begin{equation}
{\cal L}_5 = - {\kappa_{ij} \over  \Lambda} (\nu_i \phi^0 - l_i \phi^+) 
(\nu_j \phi^0 - l_j \phi^+) + H.c.,
\end{equation}
where $(\nu_i,l_i), i=1,2,3$ are the three left-handed lepton doublets of 
the SM and $(\phi^+,\phi^0)$ is the one Higgs scalar 
doublet.  As $\phi^0$ acquires a nonzero vacuum expectation value 
$\langle \phi^0 \rangle = v/\sqrt{2}=174~{\rm GeV}$, the neutrino mass matrix is given by
\begin{equation}
{\cal M}^\nu_{ij} = {\kappa_{ij} v^2 \over \Lambda}.
\end{equation}
Tree-level~\cite{Minkowski:1977sc,Mohapatra:1979ia,Yanagida:1979as,Schechter:1980gr,Foot:1988aq} and one-loop realizations~\cite{Ma:1998dn,Ma:2006km,Barbieri:2006dq, Cao:2007rm, Cao:2009yy} of this operator have 
been discussed extensively in the literature, as well as some two-loop~\cite{Cheng:1980qt,Petcov:1984nz,Zee:1985id,Babu:1988ki,Babu:1988ig,Ma:2007gq,Guo:2012ne,Angel:2012ug,Sierra:2014rxa,Farzan:2014aca} 
and three-loop~\cite{Krauss:2002px,Aoki:2008av,Gustafsson:2012vj} examples. The two-loop case is particularly 
inviting because the smallness of the neutrino mass, i.e. of order 0.1 eV, 
agrees well with a mass scale of $\mathcal{O}(1)~{\rm TeV}$ for the heavy particles in 
the loops, without unduly small and large Yukawa and scalar couplings, as would be the case with one-loop and three-loop realizations.

\begin{figure}[b]
\centering
\includegraphics[scale=0.45]{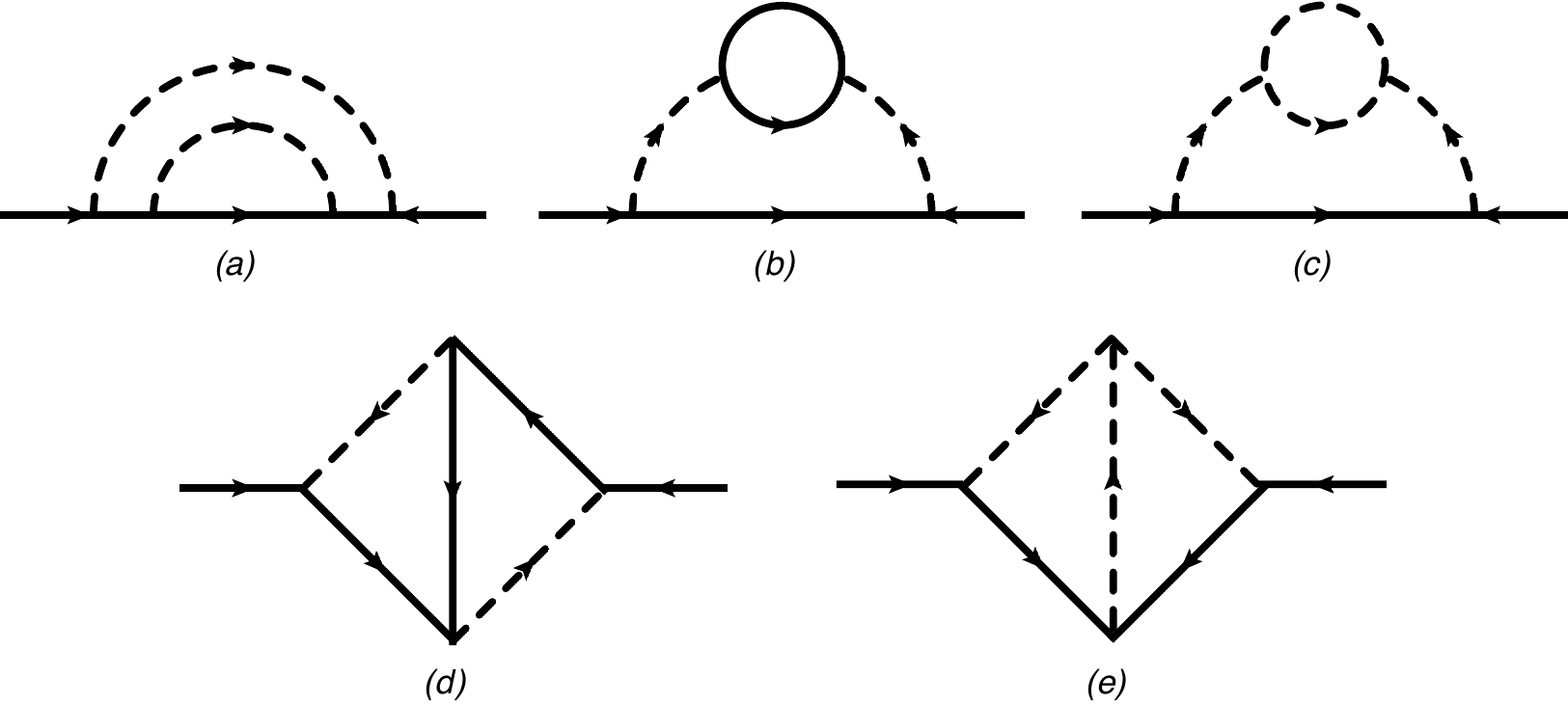}
\caption{Two-loop skeleton diagrams for neutrino mass before the Higgs insertions.}
\label{fig:2loop0}
\end{figure}

Generic structures of the two-loop diagrams involving fermions and scalars 
are shown in Fig.~\ref{fig:2loop0}, where the external Higgs lines are yet to be inserted~\cite{Sierra:2014rxa}.
To minimize the particle content inside the loops, we assume that the topology of each diagram exhibits a left-right or central symmetry. Under such a condition, Figs.~\ref{fig:2loop0}(a)-(d) may very well have a singlet Majorana fermion that could itself generate a neutrino mass at tree level.  The only exception is Fig.~\ref{fig:2loop0}(e) which is thus a truly two-loop effect.  In the following, we will focus on this case and examine the different Higgs insertions.

\begin{figure}[!h]
\centering
\includegraphics[scale=0.25]{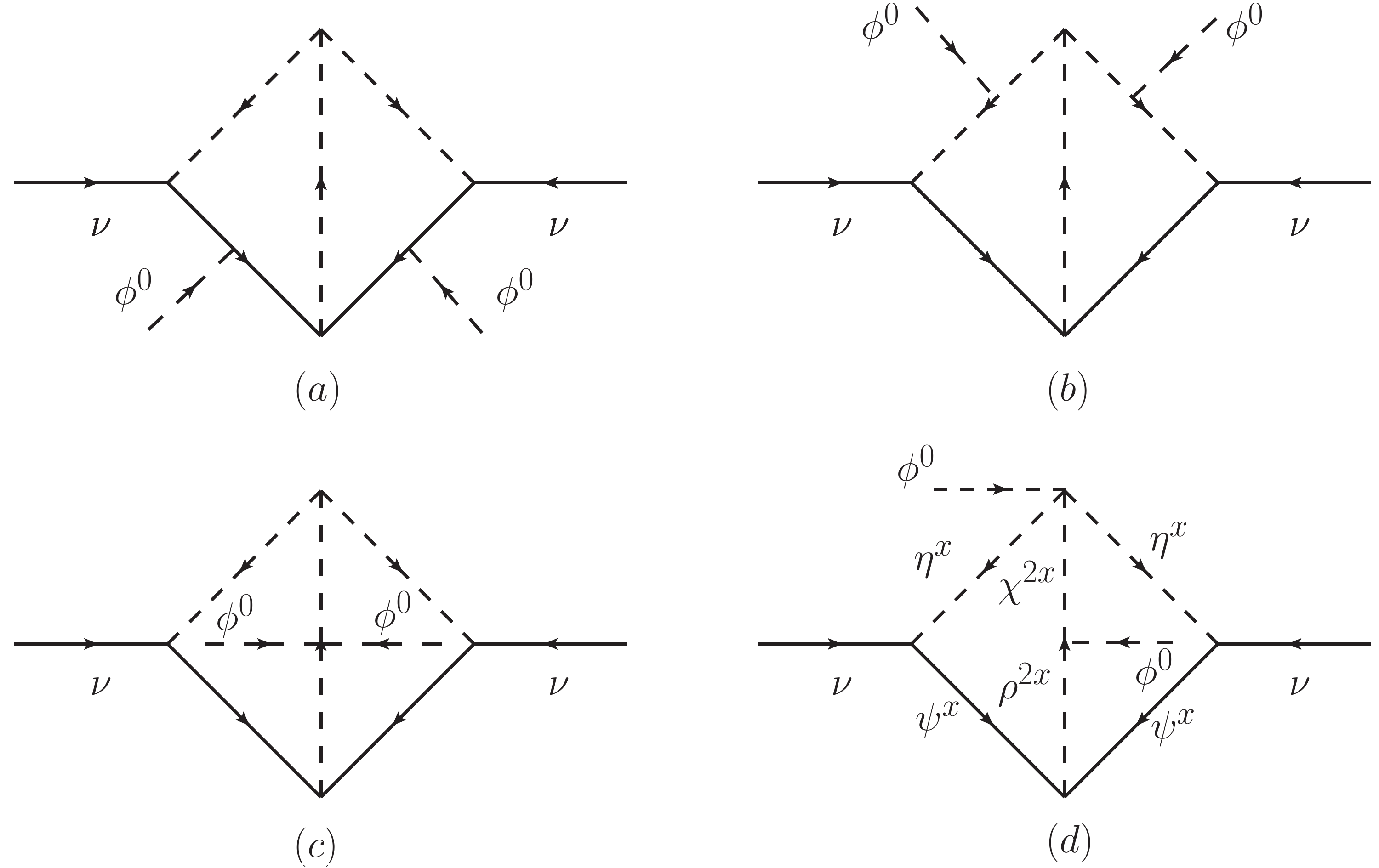}
\caption{Two-loop diagrams for neutrino mass with Higgs insertions onto 
fermion lines (a), scalar lines (b), converging on the center scalar line connecting to neutrinos (c) and one Higgs line at the top and one Higgs line at the center (d). }
\label{fig:2loop}
\end{figure}

One obvious way of inserting two external Higgs lines is onto the 
two fermion lines connecting to the external neutrino lines as shown in 
Fig.~\ref{fig:2loop}(a).  Another is onto the corresponding scalar lines as shown in Fig.~\ref{fig:2loop}(b). 
A third way also exists with the two Higgs lines converging onto the center scalar line as shown in Fig.~\ref{fig:2loop}(c).  Specific realizations of all 
these have been studied previously.  In this paper we consider for the first time specific examples of the fourth option~\cite{Angel:2012ug} with one Higgs line at the top and one Higgs line at the center as shown in Fig.~\ref{fig:2loop}(d).

We show that there are some discriminative collider signatures among these four topologies under the simplest representations of the new particles in the loop due to the different Higgs insertions. Such collider signatures may then be used to distinguish our two-loop topology (Fig.~\ref{fig:2loop}(d)) with others (Figs.~\ref{fig:2loop}(a, b, c)).

The electric charge assignments of fermion 
$\psi^x$, scalar doublets $(\eta^{x+1},\eta^x)$, $(\chi^{2x+1},\chi^{2x})$, 
and singlet $\rho^{2x}$ are as shown in Fig.~\ref{fig:2loop}(d). 
There are at least three natural realizations of this diagram: (A) $x=-1$ 
with $\psi_R = l_R$; (B) $x=-1$ with $\psi_R = E_R$ which is a new heavy fermion with a Dirac partner $E_L$ such that both $E$ and $\eta$ are odd under a dark $Z_2$ symmetry; (C) $x=-1/3$ with $\psi_R = d_R$. The quantum numbers of 
the particles in the loop under $SU(3)_C\otimes SU(2)_L\otimes U(1)_Y$ are collected in Table~\ref{tbl:models}.
It is obvious that the doubly charged scalar is predicted in (A) and (B), while a dark matter candidate is embedded in (B). 
The leptoquark and diquark scalars are related to the neutrino mass generation in (C).
It is easy to generate the neutrino mass around $0.1~{\rm eV}$ when the new particles are $\mathcal{O}(1)~{\rm TeV}$, without unduly small or large Yukawa and scalar couplings in these models. The effective Lagrangian related to our study is 
\begin{align}
\mathcal{L}_{\nu}&\supset f_{ij}\bar{\nu}_i\eta^{-x}\psi^x_{Rj}+h.c.\nn\\&+h_{ij}\overline{(\psi^x_{Ri})^c}\psi^x_{Rj}\rho^{{-2x}} +\dfrac{\lambda v}{\sqrt{2}}\eta^x\eta^x\chi^{-2x},
\end{align}
where $i,j=1,2,3$ is the family index. 
 Details of the neutrino mass generation are shown in the Appendix.  

Next we perform a collider simulation to explore the potential of the Large Hadron Collider (LHC) on discovering the three models. We focus on a 14~TeV LHC with an integrated luminosity ($\mathcal{L}$) of $100~{\rm fb}^{-1}$ and also the high-luminosity (HL) phase of $3000~{\rm fb}^{-1}$ in this study, and for simplicity, we assume the neutrino mass matrix is diagonal for the collider phenomenology analysis. 

\begin{table}
\renewcommand{\arraystretch}{1.5}
\begin{tabular}{|c|c|c|c|c| }\hline
Model & $\eta$ & $\chi$ & $\rho$ & $\psi$ \\
\hline
(A) & (1, 2, $-\frac{1}{2}$) & (1, 2, $\frac{3}{2}$) & (1, 1, 2) & $(1, 1, -1)$ \\
\hline
(B) & (1, 2, $-\frac{1}{2}$, $-$) & (1, 2, $\frac{3}{2}$, +) & (1, 1, 2, +) & $(1, 1, -1, -)$ \\
\hline
(C) & (3, 2, $\frac{1}{6}$) & ($\bar{6}$, 2, $\frac{1}{6}$) & ($\bar{6}$, 1, $\frac{2}{3}$) & (3, 1, $-\frac{1}{3}$) \\
\hline
\end{tabular}
\caption{Loop particles under $SU(3)_C\otimes SU(2)_L\otimes U(1)_Y\otimes Z_2$ of the three models, where $Z_2$ applies only to (B) and $\psi$ denotes 
$l_R$, $E_R$, $d_R$ in (A), (B) and (C), respectively.
\label{tbl:models}
}
\end{table}

~\\
\noindent{\bf Collider Phenomenology of Model (A).}

\noindent{\it Discovery potential.}  The best way to probe our model is the pair production of the doubly charged scalar ($\chi^{++}$)
\beq
pp\to \chi^{++} \chi^{--},~\chi^{++}\to \ell^+\ell^+,~\chi^{--}\to \ell^-\ell^-, \nn
\eeq
where $\ell^{\pm}$ represents the electron ($e^{\pm}$) and muon ($\mu^{\pm}$). The event topology is characterized by four isolated charged leptons. The dominant backgrounds are $\gamma\gamma$, $Z\gamma$, $ZZ$, four leptons ($4\ell$), and four charged leptons plus one jet ($4\ell 1j$). The null result in the search of doubly charged scalars via the pair production at the 8~TeV LHC imposes a bound, $m_{\chi^{++}}>400~{\rm GeV}$, 
assuming $\chi^{++}$ decays entirely into electron or muon pairs~\cite{ATLAS:2014kca, CMS:2016cpz}. 

We generate both the signal and the background processes at the parton level using MadEvent~\cite{Alwall:2011uj} and impose basic cuts as follows: $p_T^{\ell^{\pm},j}>5~{\rm GeV}$ with $\left|\eta^{\ell^\pm,j}\right|<5$, where $p_T$ and $\eta$ denote the transverse momentum and rapidity, respectively. We require the angular distance $\Delta R_{mn}\equiv\sqrt{(\eta^m-\eta^n)^2+(\phi^m-\phi^n)^2}$ between the objects $m$ and $n$ to be greater than 0.4 to obtain isolated objects. At the analysis level, all the signal and background events are required to pass a set of selection cuts~\cite{CMS:2012kua,ATLAS:2014kca}:
\bea
&&p_T^{\ell_1}>20~{\rm GeV}, \quad  p_T^{\ell_2}>15 ~{\rm GeV}, \quad p_T^{\ell_{3,4}}>10 ~{\rm GeV},\nn\\
&&|\eta^{\ell}|\leq 2.5, \quad\quad\quad \met<40~{\rm GeV},
\label{eq:cut}
\eea
where $\ell_i$ with $i=1,2,3,4$ denotes the lepton ordered in accord with their $p_T^{\ell}$'s. We demand only four leptons in the central region of the detector and veto extra jets if $p_T^j>10~{\rm GeV}$ or $|\eta^j|>3.5$.
In addition, we require the invariant mass of lepton pair to be away from $m_Z$, with $|m(\ell\ell^\prime)-m_{Z}|>10 ~{\rm GeV}$, to suppress the dominant background containing  $Z$-boson resonances. In order to suppress the $4\ell1j$ background, we require $m(\ell^{\pm}\ell^{\mp})>50~{\rm GeV}$. We end up with 26.05 background events in total at the LHC with $\mathcal{L}=100~{\rm fb}^{-1}$, i.e. $\gamma\gamma~(1.18)$, $Z\gamma~(2.67)$, $ZZ~(3.62)$, $4\ell~(12.03)$ and $4\ell1j~(6.56)$. The number inside the parenthesis denotes the number of events of each individual background.

\begin{figure}
\includegraphics[scale=0.3]{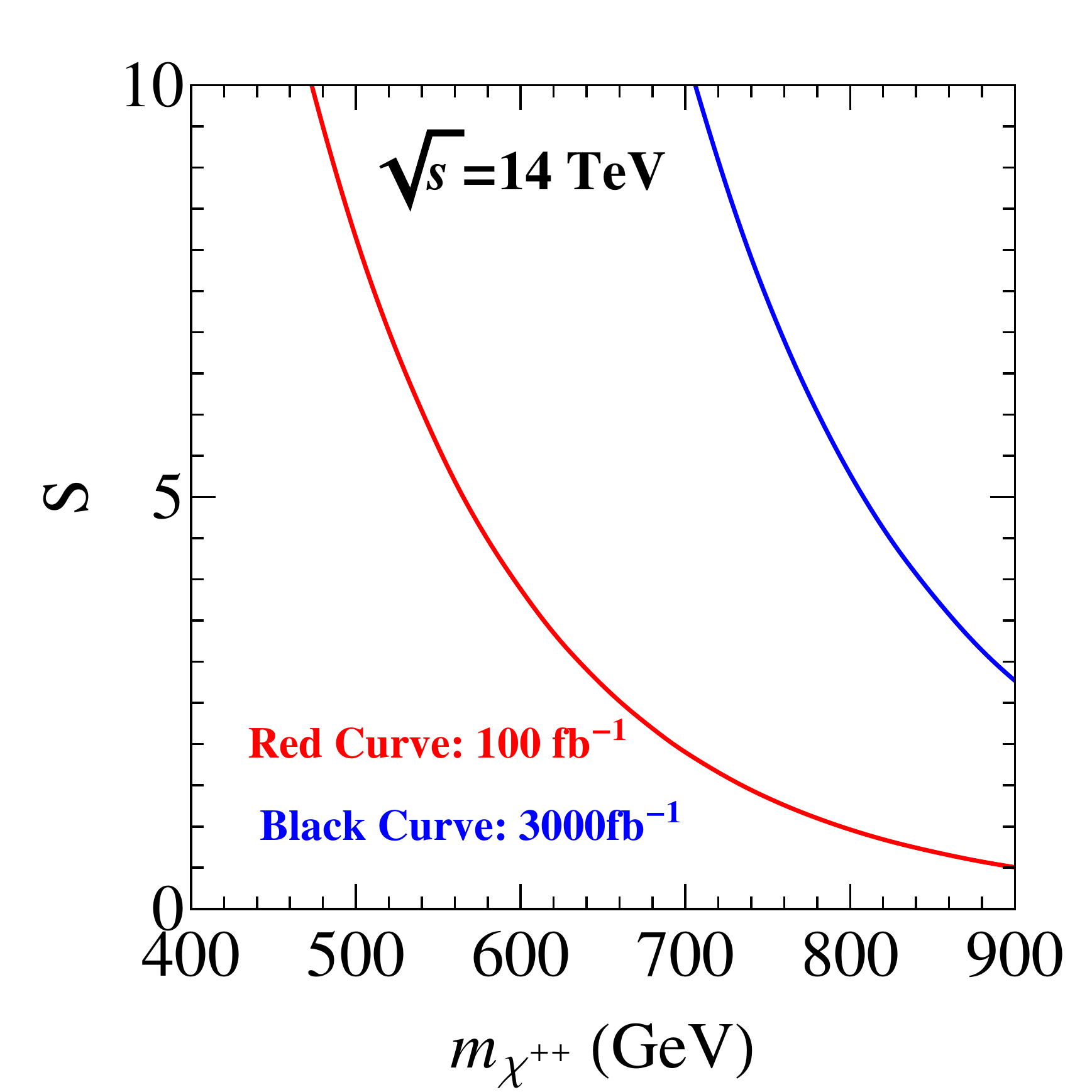}
\caption{The significance of discovering $\chi^{++}$ at the LHC with $\mathcal{L}=100~{\rm fb}^{-1}$ (red line) and HL-LHC (blue line).}
\label{fig:siga0}
\end{figure}

We obtain a 5 standard deviations ($\sigma$) statistical significance using
\bea
S=\sqrt{-2\left[(n_b+n_s)\log\frac{n_b}{n_s+n_b}+n_s\right]}=5,
\eea
where $n_b$ and $n_s$ represent the numbers of the signal and background events, respectively. Figure \ref{fig:siga0} displays the discovery potential of $\chi^{++}$ with ${\rm BR}(\chi^{++}\to \ell^+\ell^+)=1$ at the LHC. The $\chi^{++}$ with $m_{\chi^{++}}<566~{\rm GeV}$ could be discovered with $\mathcal{L}=100~{\rm fb}^{-1}$ (red line). The HL-LHC extends the coverage to  $m_{\chi^{++}}<806~{\rm GeV}$ (blue line).

~\\
\noindent{\it Model Discrimination.} The doubly charged scalar appears in all the two-loop models depicted in Fig.~\ref{fig:2loop} and yields exactly the same collider signature. Observing a doubly charged scalar alone cannot distinguish various models. However, different insertions of Higgs fields in Fig.~\ref{fig:2loop} offer an opportunity to distinguish our two-loop topology with others if new particles inside the loops are in the simplest representations under the SM gauge group. Our model consists of two special ingredients: 
\begin{itemize}
\item[(i)] two doublet scalars $\chi$ and $\eta$; 
\item [(ii)] the quartic coupling $(\chi\epsilon\eta) (\tilde{\phi}\epsilon\eta)$. 
\end{itemize}
The scalars in other two-loop models exhibit different weak quantum numbers. For example, the singly charged scalar $(\kappa^+)$ and doubly charged scalar $(\chi^{++})$ in Fig.~\ref{fig:2loop}(a) are both neutral under $SU(2)_L$; a good example is the so-called Babu-Zee model~\cite{Zee:1985id,Babu:1988ki}. Fig.~\ref{fig:2loop}(b) consists of two singlet scalars and one doublet scalar with hypercharge 1/2~\cite{Okada:2014qsa}. Fig.~\ref{fig:2loop}(c) is used in Ref.~\cite{Guo:2012ne} to discuss neutrino mass generation which involves a triplet and two singlet scalars.

The simplest way to distinguish our model shown in Fig.~\ref{fig:2loop}(d) from the models in Figs.~\ref{fig:2loop}(a) and (b) is the $\chi^{++}$ and $\chi^-$ associated production, i.e. 
\beq
pp\to \chi^{\pm\pm}\chi^{\mp},\quad \chi^{\mp} \to \chi^{\mp\mp(\star)}W^\pm.\nn
\eeq
This process is absent in the models shown in Fig.~\ref{fig:2loop}(a) and Fig.~\ref{fig:2loop}(b). The $\chi^{--}$ can be produced on-shell or off-shell (labeled as $^\star$), depending on  whether $\chi^+$ is heavier than $\chi^{++}$ or not. 

Note that the $\chi^\pm$ scalar can also be probed in the $\chi^+\chi^-$ pair production that yields a signature of multiple charged leptons and missing transverse momentum. However, in the parameter space of interest to us, $m_{\chi}>250 ~{\rm GeV}$, the cross section of $pp\to \chi^+\chi^-$ with subsequent decays $\chi^\pm\to \chi^{\pm\pm}W^\mp\to \ell^\pm\ell^\pm\ell^\mp+\met$ is around $10^{-4}\sim 10^{-2}~{\rm fb}$  at the 8~TeV LHC while fixing $m_{\chi^{++}}=400~{\rm GeV}$. Such a small cross section is consistent with current experimental bounds~\cite{Aad:2014nua}.

\begin{figure}
\includegraphics[scale=0.3]{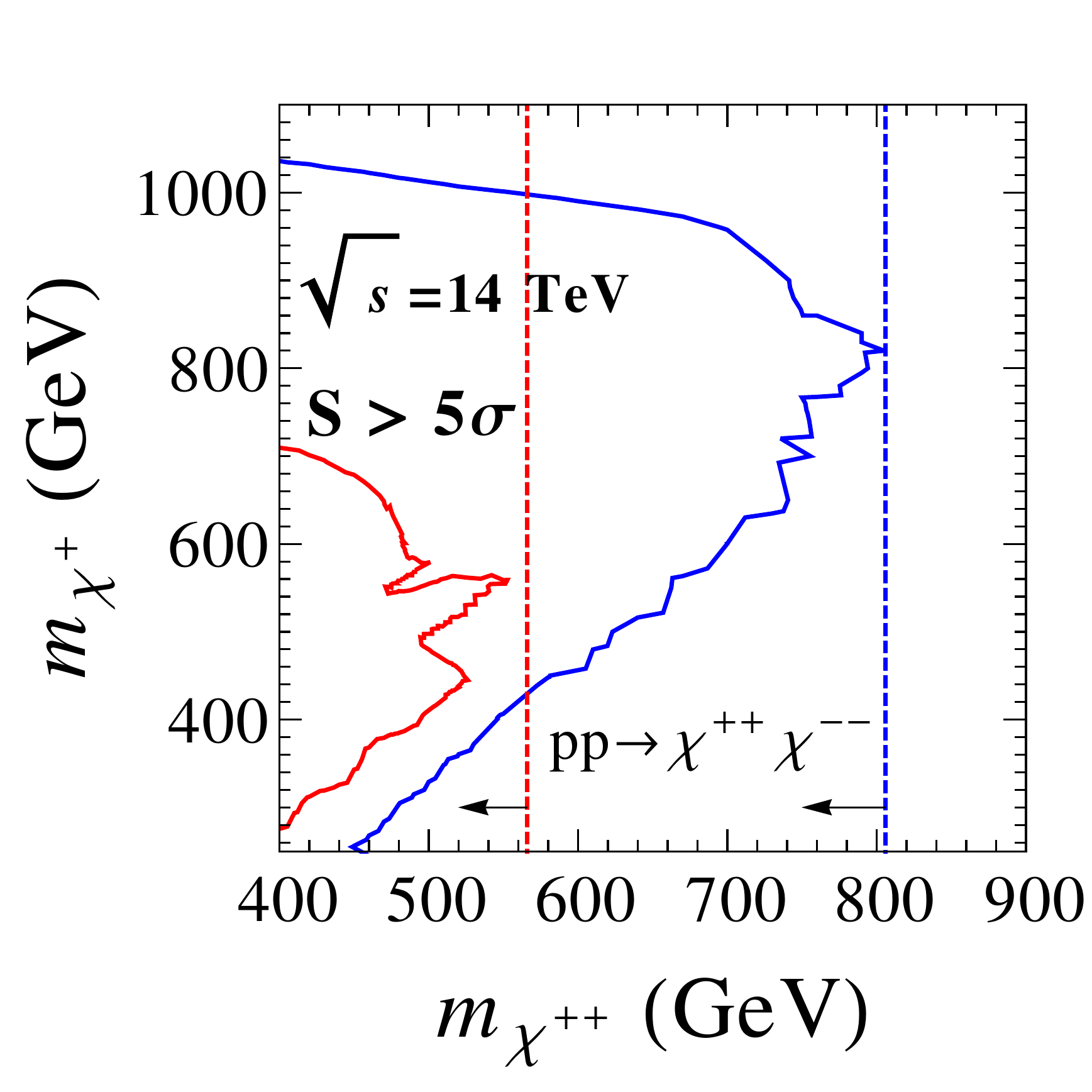}
\caption{The $5\sigma$ discovery region of  $pp\to \chi^{\pm\pm}\chi^\mp$ at the 14~TeV LHC with $\mathcal{L}=100~{\rm fb}^{-1}$ (red) and HL-LHC (blue), respectively, assuming ${\rm BR}(\chi^{++}\to \ell^+\ell^+)=1$. The dashed vertical lines denote the $5\sigma$ discovery region obtained from doubly charged scalar pair production. }
\label{fig:siga}
\end{figure}

The typical cross section of the process $pp\to \chi^{\pm\pm}\chi^{\mp}\to \chi^{\pm\pm}\chi^{\mp\mp(\star)}W^\pm$ at the 14~TeV LHC is around $0.1\sim 1 ~{\rm fb}$. Hence the hadronic modes of the $W$ boson, which exhibit large branching ratios, are considered in our simulation. For simplification, we assume $\chi^{\pm\pm}$ decays into same sign lepton pairs entirely, i.e. ${\rm BR}(\chi^{++}\to \ell^+\ell^+)=1$. To mimic the signal events, the SM backgrounds should consist of $W/Z/\gamma$ in final state. We consider the following backgrounds: $4\ell2j$,   $t\bar{t}Z$, $Z\gamma jj$, $\gamma\gamma jj$ and $t\bar{t}\gamma$. The other backgrounds like $WZZ$ and $h(\to ZZ^\star)jj$ are negligible after imposing kinematic cuts. At the analysis level, we demand only four charged leptons and two jets in the central region of the detector and require the same kinematic cuts on charged leptons as we did in the analysis of $\chi^{++}\chi^{--}$ pair production; see Eq.~\ref{eq:cut}. For the jets, we also require $p_T^j>20~{\rm GeV}$ and $|\eta^{j}|\leq 2.5$. We end up with 3.94 background events in total with $\mathcal{L}=100~{\rm fb}^{-1}$, i.e. $4\ell2j~(1.94)$, $t\bar{t}Z~(0.46)$, $Z\gamma jj~(1.03)$, $\gamma\gamma jj~(0.21)$ and $t\bar{t}\gamma~(0.30)$. 

We plot the $5\sigma$ discovery potential of $pp\to\chi^{++}\chi^{-}$ at the LHC in Fig.~\ref{fig:siga}(a), where the dashed vertical lines represent the discovery potential obtained from the doubly charged scalar pair production. Two benchmark integrated luminosities, $\mathcal{L}=100~{\rm fb}^{-1}$ (red) and HL-LHC (blue), are considered. 
It shows that $\chi^{\pm\pm}\chi^\mp$ pair production could be discovered when $m_{\chi^{++}}<550~{\rm GeV}$ and $m_{\chi^+}\sim 400-600~{\rm GeV}$ with $\mathcal{L}=100~{\rm fb}^{-1}$; see the red contour. The parameter space extends to $m_{\chi^{++},\chi^{+}}\lesssim 800~{\rm GeV}$ at the HL-LHC. Hence, we are able to discriminate between our model shown in Fig.~\ref{fig:2loop}(d) and those models in Figs.~\ref{fig:2loop}(a, b).

Note that the triplet scalar in the model of Fig.~\ref{fig:2loop}(c) can also generate the $pp\to\chi^{\pm\pm}\chi^{\mp}$  collider signature.  To distinguish our model from it, we make use of the quartic coupling $(\chi\epsilon\eta) (\tilde{\phi}\epsilon\eta)$; for example, it gives rise to unique decay modes of $\chi^{++} $ and $\chi^-$ as follows:
\bea
\chi^{++} &\to& h \eta^+ \eta^+,~h\to b \bar{b},~\eta^+\to \ell^+ \nu_l, \eta^+\to \ell^+ \nu_l,\nn\\
\chi^- &\to& h \eta^- \eta^0,~h\to b\bar{b},~\eta^-\to \ell^- \nu_l, \eta^0\to \ell^+ \ell^-,\nn
\eea
which cannot occur in Figs.~\ref{fig:2loop}(a, b, c).
The detailed analysis of the unique modes of $\chi^{\pm\pm}$ and $\chi^\pm$ mentioned above is beyond the scope of the current paper and will be presented elsewhere.

~\\
\noindent{\it Comment on the $T$ parameter.}

\begin{figure}
\includegraphics[scale=0.24]{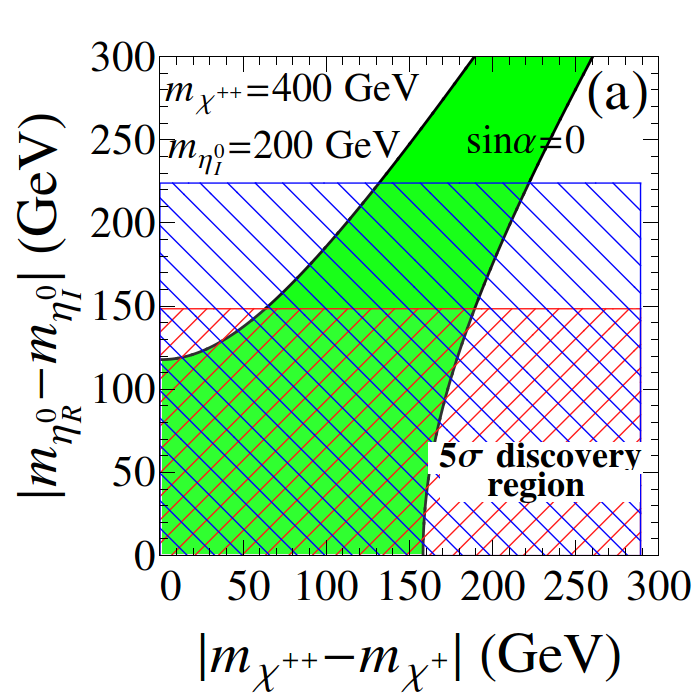}
\includegraphics[scale=0.24]{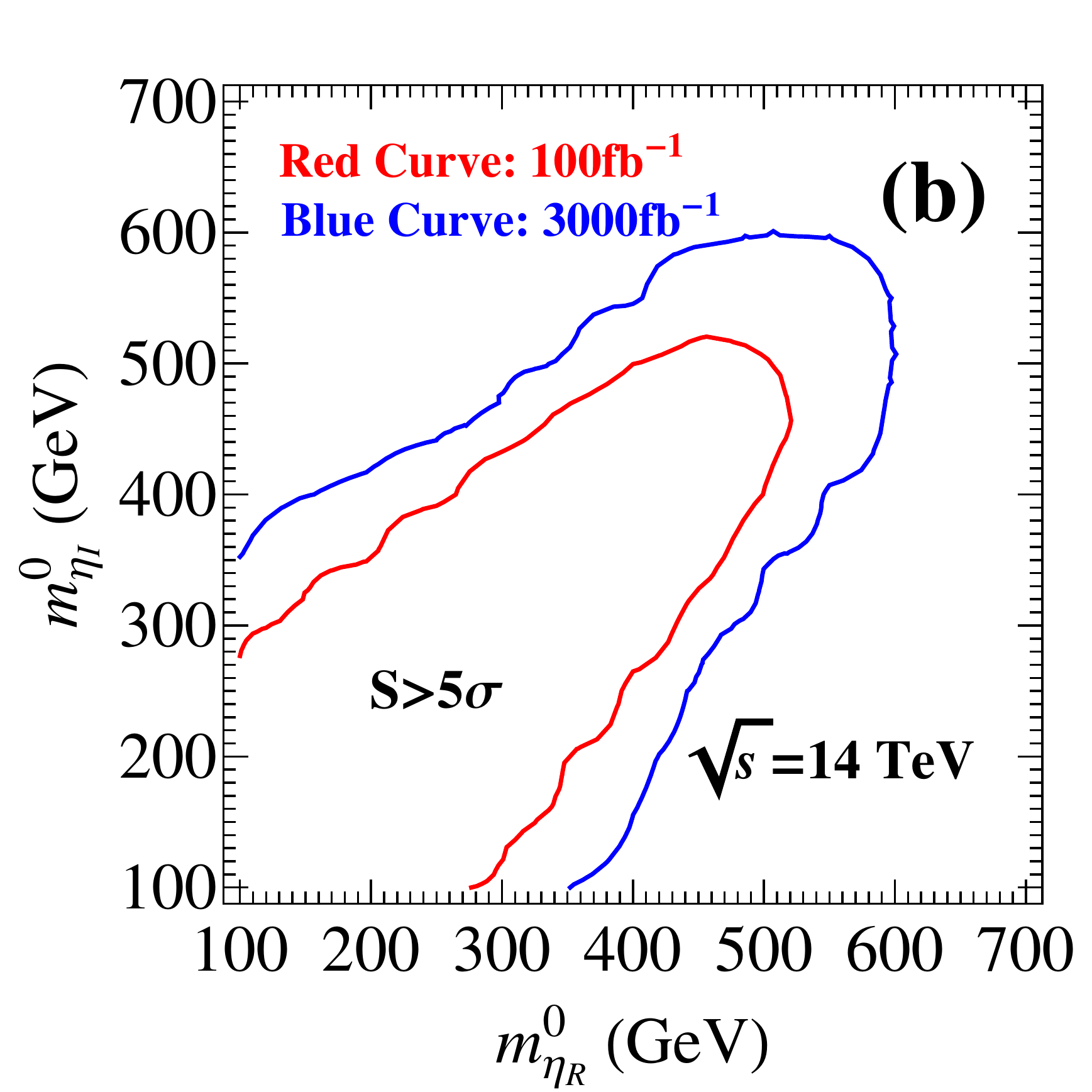}
\caption{(a) Parameter space allowed by the $T$ parameter in the plane of $|m_{\chi^{++}}-m_{\chi^+}|$ and $|m_{\eta^0_R}-m_{\eta^0_I}|$  (green region) for the parameter choice specified in the text, where the meshed region denotes the $5\sigma$ discovery region covered by the $\chi^{\pm\pm}\chi^\mp$ and $\eta^0_R\eta^0_I$ productions at the LHC with $\mathcal{L}=100~{\rm fb}^{-1}$ (red) and HL-LHC (blue), respectively. (b) The $5\sigma$ discovery region of  $pp\to Z\to \eta^0_R\eta^0_I$ with ${\rm BR}(\eta^0_{R/I}\to \ell^+\ell^+)=1$ at the LHC with $\mathcal{L}=100~{\rm fb}^{-1}$ (red) and HL-LHC (blue), respectively. }
\label{fig:sigaT}
\end{figure}

It is well known that the triplet scalar violates the condition for the $\rho$ parameter to be one if it develops a vacuum expectation value. That restricts the scalar potential of the model shown in Fig.~\ref{fig:2loop}(c). On the other hand, the scalar doublet $\eta=(\eta^0,\eta^-)$ in our model, where $\eta^0=(\eta^0_R+i\eta^0_I)/\sqrt{2}$, also contributes to the $T$ parameter (which is proportional to $\rho-1$) and relaxes the constraint on the mass splitting between $\chi^{++}$ and $\chi^+$. 

The $T$ parameter is not only sensitive to the mass splitting between $\eta^-$ and $\eta^0_{R,I}$, but also to the difference of $m_{\eta^0_{R}}$ and $m_{\eta^0_I}$~\cite{Okada:2014qsa}. Relaxing the $T$ parameter constraint imposes a strong correlation between $(m_{\chi^{++}}-m_{\chi^+})$ and $(m_{\eta^0_R}-m_{\eta^0_I})$. Figure~\ref{fig:sigaT}(a) shows the allowed parameter space by the $T$ parameter constraint at the 95\% confidence level for the benchmark parameters of $\sin\alpha=0$, $m_{\eta^0_I}=200~{\rm GeV}$, $m_{\chi^{++}}=400~{\rm GeV}$ and $m_{\eta^+}=(m_{\eta^0_R}+m_{\eta^0_I})/2$; see the green region.  Here, $\alpha$ is the mixing angle between the two doubly charged scalars in our model. One can safely ignore it in the numerical evaluation of the $T$ parameter although the neutrino mass demands a tiny nonzero $\sin\alpha$.

For completeness, we also investigate the potential of the LHC on discovering the neutral scalars $\eta^0_{R}$ and $\eta^0_{I}$ through the process of $pp\to Z\to\eta^0_R(\to\ell^+\ell^-)\eta^0_I(\to\ell^+\ell^-)$.  This channel gives rise to exactly the same collider signature as the $\chi^{++}\chi^{--}$ pair production, and we show the  $5\sigma$ discovery potential of $\eta^0_R\eta^0_{I}$ pairs in Fig.~\ref{fig:sigaT}(b). 

The meshed regions in Fig.~\ref{fig:sigaT}(a) denote the $5\sigma$ discovery regions of the processes $pp\to\chi^{\pm\pm}\chi^\mp$ and $pp\to\eta^0_R\eta^0_I$ at the LHC. They show that the LHC with $\mathcal{L}=100~{\rm fb}^{-1}$ could probe the mass splitting between $\chi^{++}$ and $\chi^+$ up to 170 GeV maximally as well as the mass splitting of $\eta^0_R$ and $\eta^0_I$ up to 150 GeV (red region). The HL-LHC would extend the reach in both mass splittings to 224 GeV (blue region).

~\\
\\
\noindent{\bf Collider Phenomenology of Models (B) and (C).}

In (B), the vector-like charged fermion $E^+$ is odd under dark $Z_2$ and is connected to the dark matter candidate, i.e. $\eta^0_R$ or $\eta^0_I$ whichever is lighter.  The best way to probe the dark fermion $E^+$ is through $pp\to E^+E^-\to\ell^+\ell^-\eta^0_{R/I}\eta^0_{R/I}$.   It yields a collider signature of two charged leptons plus missing energy, which is shared by many neutrino mass models~\cite{Ding:2016wbd,Guo:2016dzl,vonderPahlen:2016cbw}.
The dark matter relic abundance requires $m_{\eta^0_{R/I}}\sim 63~{\rm GeV}$~\cite{Okada:2014qsa}. The mass of $E^+$ is severely constrained by the slepton search $pp\to\tilde{\ell}^+\tilde{\ell}^-\to \ell^+\ell^-\tilde{\chi}^0\tilde{\chi}^0$~\cite{Aad:2014vma,Khachatryan:2014qwa}, where $\tilde{\ell}^\pm$ and $\tilde{\chi}^0$ are the charged slepton and the neutralino, respectively. Suppose ${\rm BR}(E^+\to\ell^+\eta^0_{R/I})=1$ and $m_{\eta^0_{R/I}}=63~{\rm GeV}$, we obtain $m_{E^+}>400~{\rm GeV}$. 

\begin{figure}[b]
\includegraphics[scale=0.24]{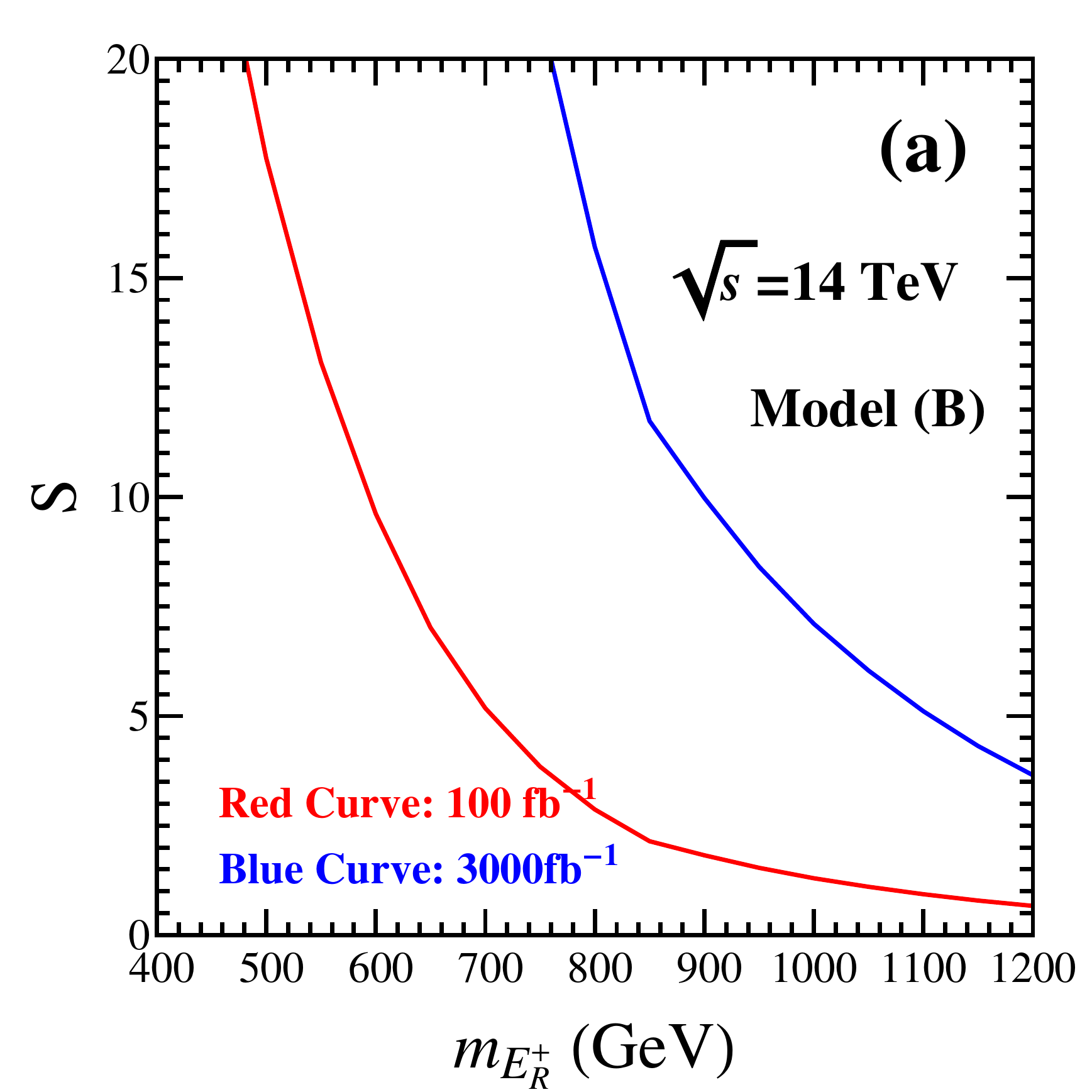}
\includegraphics[scale=0.24]{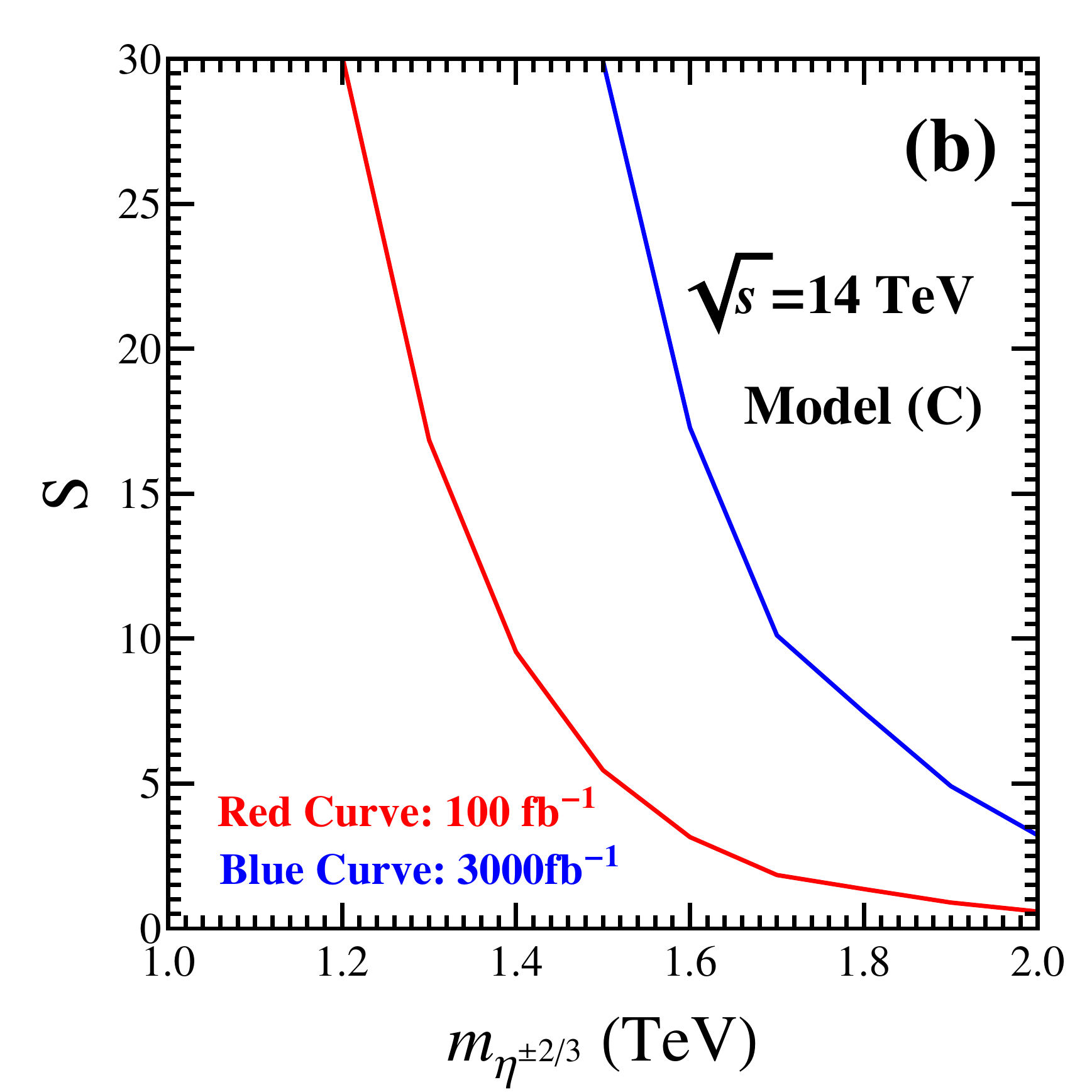}
\caption{(a) Discovery potential of $E^+$ in (B) with $m_{\eta^0_{R/I}}=65~{\rm GeV}$ and ${\rm BR}(E^+\to\ell^+\eta^0_{R/I})=1$  at the LHC with $\mathcal{L}=100~{\rm fb}^{-1}$ (red) and HL-LHC (blue); (b) Discovery potential of leptoquark $\eta^{\pm 2/3}$ in (C) with ${\rm BR}(\eta^{\pm 2/3}\to \ell^{\pm}j)=1$.}
\label{fig:sigER}
\end{figure}

We follow the slepton search at the 8~TeV LHC~\cite{Aad:2014vma,Khachatryan:2014qwa} to perform a simulation to estimate the potential of observing $E^+E^-$ pairs at the 14~TeV LHC with $\mathcal{L}=100~{\rm fb}^{-1}$. The dominant backgrounds include the diboson productions ($WW$, $WZ$ and $W\gamma$), triple-boson productions ($WWZ$ and $WWW$), and $tW$ production. 
We demand two isolated charged leptons with $p_T^{\ell(1)}>150~{\rm GeV}$, $p_T^{\ell(2)}>100~{\rm GeV}$ and $|\eta^{\ell}|<2.5$, and veto extra jets if $p_T^j>10~{\rm GeV}$ or $|\eta^j|>3.5$. To suppress the backgrounds, we also require $\met>200~{\rm GeV}$ and $m(\ell^+\ell^-)>200~{\rm GeV}$. We end up with 35.1 background events, i.e. $WW~(31.2)$, $WZ~(0.2)$, $W\gamma~(0.2)$, $tW~(2.6)$, $WWW~(0.9)$ and $WWW~(0.04)$, where the number inside the parenthesis denote the number of events of the corresponding channel. That yields a discovery potential of $E^{\pm}$ with ${\rm BR}(E^+\to\ell^+\eta^0_{R/I})=1$ depicted in Fig.~\ref{fig:sigER}(a), which shows $E^+$ could be discovered when $m_{E^+}<700~{\rm GeV} $ with $100~{\rm fb}^{-1}$ (red). The HL-LHC would extend the coverage to $m_{E^+}<1105~{\rm GeV}$ (blue).

In (C), the neutrino mass is generated through $SU(3)_C$ colored particles, i.e. leptoquark and diquark scalars which also give rise to a very rich phenomenology. The current constraint of the first and second generation leptoquark is $m_{\eta}> 1~{\rm TeV}$~\cite{Aad:2015caa,Aaboud:2016qeg,CMS:2016qhm}. The process $pp\to\eta^{2/3}\eta^{-2/3}$ with subsequent decays of $\eta^{\pm 2/3}\to\ell^{\pm} j$ can efficiently probe such a leptoquark and has been widely studied in the literature~\cite{Mandal:2015lca,Dorsner:2016wpm}. The diquark is severely constrained by the dijet data at the 13~TeV LHC~\cite{Khachatryan:2015dcf}. However, the lower limit of the diquark mass depends critically on the Yukawa coupling $h_{ij}\overline{q^c_{Ri}}q_{Rj}\chi$. To generate the tiny neutrino mass, we choose $h_{ij}\sim\mathcal{O}(0.01)$ (see the appendix), which yields the cross section of $pp\to \chi^{\pm2/3}$ around $1~{\rm pb}$ for $m_{\chi^{\pm 2/3}}=400~{\rm GeV}$ at the 13 TeV LHC. That is much smaller than the current experimental constraint~\cite{Khachatryan:2015dcf}. A much stronger constraint is obtained from the search of R-parity violating decays of the top squark in four-jet final states ($pp\to\tilde{t}\tilde{t}\to 4j$) at the 13~TeV LHC~\cite{ATLAS:2016yhq}. We find $m_{\chi^{\pm 2/3}}$ is larger than $700 \sim 800$ GeV with the assumption of ${\rm BR}(\chi^{\pm 2/3}\to jj)=1$. This bound does not rely on the Yukawa coupling. 

The best way to verify (C) is through the leptoquark $\eta^{\pm 2/3}$ pair production at the LHC. The main SM backgrounds are $t\bar{t}~ (2.8)$, $t\bar{t}j~(1.3)$ and $WWjj~(11.1)$, where the number inside the parenthesis denotes the number of events of corresponding channel at the LHC with $\mathcal{L}=100~{\rm fb}^{-1}$ after cuts shown below: (i) exactly two lepton and two jets with $p_T^{\ell, j(1)}>300~{\rm GeV}$, $p_T^{\ell, j(2)}>200~{\rm GeV}$ and $|\eta^{\ell,j}|<2.5$; (ii) $\met<40~{\rm GeV}$. The discovery potential of leptoquark with ${\rm BR}(\eta^{\pm 2/3}\to \ell^{\pm}j)=1$ is shown in Fig.~\ref{fig:sigER}(b). 
It shows that $\eta^{\pm 2/3}$ could be discovered for $m_{\eta^{\pm 2/3}}<1.5~{\rm TeV}$ (red). The HL-LHC extends the coverage to $m_{\eta^{\pm 2/3}}<1.9~{\rm TeV}$ (blue).
 
\noindent{\bf Summary.}
In this work, we have constructed a new class of two-loop models of neutrino mass where the two external Higgs lines are attached differently from previous examples. There are three natural realizations containing (A) doubly charged scalar, (B) dark matter candidate, and (C) leptoquark and diquark scalars. We show that these new particles have promising discovery potentials at the $\sqrt{s}=14~{\rm TeV}$ LHC with an integrated luminosity $\mathcal{L}=100~{\rm fb}^{-1}$ and $3000~{\rm fb}^{-1}$.  We demonstrate as well how the different Higgs insertions offer an opportunity of distinguishing these models from others under the assumption that the new particles in the loop are in the simplest representations of the SM gauge group. 
As an example, we propose that the $pp\to \chi^{\pm\pm}\chi^{\mp}$ production process in (A) could be used to distinguish our model topology (see Fig.~\ref{fig:2loop}(d)) from Fig.~\ref{fig:2loop}(a) and  Fig.~\ref{fig:2loop}(b). It is also possible to distinguish it from Fig.~\ref{fig:2loop}(c) by using the quartic coupling $(\chi\epsilon\eta) (\tilde{\phi}\epsilon\eta)$. In addition, two doublet scalars inside the loop of our model could be useful to relax the constraint from the $T$ parameter. 
Models (B) and (C) also share the same topology with (A) and the same promise to be probed at the LHC, with possible signatures to distinguish their different Higgs insertions.

This work is supported in part by the National Science Foundation of China (11175069, 11275009, 11422545, 11675002, 11635001) and the U.~S.~Department of Energy under Grant No.~DE-SC0008541.

~\\

\noindent{\bf Appendix: Neutrino Mass.} 
In (A) and (C), the fermions in the loop are the SM right-handed leptons or quarks. The neutrino mass matrix is
\begin{align}\label{eq:mv}
&M_{ab}=\sum_{c,d}\sqrt{2}\cos\alpha \sin\alpha f_{ac}h_{cd}^\star f_{bd}\lambda v ~\times \\
&\left[I_{acbd}(m_c,m_d,m_\eta,m_{H_2})-I_{acbd}(m_c,m_d,m_\eta,m_{H_1})\right],\nn
\end{align}
where $m_c$ and $m_d$ are the fermion masses in the loop (right handed lepton or quark in the SM).
The doublet $\chi^{2x}$ mixes with $\rho^{2x}$ through the SM Higgs doublet $\phi^0$. That yields two physical scalars, $H_1$ and $H_2$, and both scalars contribute to the neutrino mass matrix. For simplification we assume $m_{H_1}\geq m_{H_2}$. The parameter $\alpha$ is the mixing angle which diagonalizes the $\chi^{2x}$ and $\rho^{2x}$ fields. The mixing effect also leads to a cancellation between $H_1$ and $H_2$. In other words, the neutrino mass is sensitive to the mass splitting of $H_1$ and $H_2$.
Degenerate $H_1$ and $H_2$ masses will give a massless neutrino in that topology.
The loop function $I_{acbd}$ is given by
\begin{align}
&I_{acbd}(m_c,m_d,m_\eta,m_{H_i})=\dfrac{1}{(16\pi^2)^2}\times\nn\\
&\int_0^1dx\int_0^1dy
\left[\dfrac{t_1^2\log{t_1}}{(t_1-1)(t_1-t_2)}-\dfrac{t_2^2\log{t_2}}{(t_2-1)(t_1-t_2)}\right],\nn
\end{align}
where $t_1=m_c^2/M_i^2$ and $t_2=m_{\eta}^2/M_i^2$ with
\bea
M_i^2=\dfrac{(1-y)m_{H_i}^2+y(1-x)m_{\eta}^2+xym_d^2}{y(1-y)}.\nn
\eea
Note $I_{acbd}$ is symmetric about exchanging $t_1$ and $t_2$, and $M_i$ is also symmetric if we swap both $m_d\leftrightarrow m_{\eta}$ and $x\leftrightarrow (1-x)$. Such a behavior can be understood from the topology of the neutrino mass generation mechanism (see Fig.~\ref{fig:2loop} (d)). In the limit of zero neutrino momentum, $\psi^x$ and $\eta^x$ play the same role in the loop integration.

Assuming $m_c,m_d\ll m_\eta,m_{H_i}$, the loop function $I_{acbd}(m_c,m_d,m_\eta,m_{H_i})$ is simplified:
\bea
I_{acbd}\simeq \dfrac{1}{(16\pi^2)^2}\int_0^1dx\int_0^1dy\left[\dfrac{-\log\Delta_i}{1-\Delta_i}\right],\nn
\eea
where 
\bea
\Delta_i=\dfrac{(1-y)r_i+y(1-x)}{y(1-y)},\nn
\eea
with $r_i=m_{H_i}^2/m_\eta^2$.

In (B) the vector-like fermion $E$ is introduced to generate the neutrino mass. There is a term arising from $m_{E}$ in addition to that of Eq.~\ref{eq:mv}. The mass matrix, similar to that in the Babu-Zee model, is given by~\cite{Babu:2002uu,McDonald:2003zj}
\begin{align}
&{M}^\prime_{ab}=\sum_{c,d}\sqrt{2}\cos\alpha \sin\alpha f_{ac}h_{cd}^\star f_{bd}\lambda v m_cm_d\nn\\
&\times\left[\widetilde{I}_{acbd}(m_c,m_d,m_\eta,m_{H_2})-\widetilde{I}_{acbd}(m_c,m_d,m_\eta,m_{H_1})\right],
\nn
\end{align}
where the loop function $\widetilde{I}_{acbd}(m_c,m_d,m_\eta,m_{H_i})$ is
\begin{align}
&\widetilde{I}_{acbd}(m_c,m_d,m_\eta,m_{H_i})=\dfrac{1}{(16\pi^2)^2}\int_0^1dx\int_0^{1}dy\nn\\
&\left[\dfrac{t_1\log{t_1}}{(t_1-1)(t_1-t_2)}
	   -\dfrac{t_2\log{t_2}}{(t_2-1)(t_1-t_2)}\right]\dfrac{1}{M_i^2(1-y)}.\nn
\end{align}
Similar to $I_{acbd}$, the loop function $\widetilde{I}_{acbd}$ is also symmetric from exchanging $t_1$ and $t_2$.
We also note that there is a strong correlation between $\widetilde{I}_{acbd}$ and  $I_{acbd}$ due to the similar loop integration. However, $I_{acbd}$ is proportional to the integration momentum $p^2$, while $\widetilde{I}_{acbd}$ is depending on the fermion mass $m_cm_d$. That leads to the different power dependence on $t_1$ and $t_2$.
In the limit of $m_{c,d} \to 0$, $\widetilde{I}_{acbd}(m_c,m_d,m_\eta,m_{H_i})$ is given by 
\bea
\widetilde{I}_{acbd}\simeq\dfrac{1}{(16\pi^2)^2m_{\eta}^2}\int_0^1dx\int_0^{1}dy\left[\dfrac{-\log\Delta_i}{(1-\Delta_i)}\dfrac{1}{1-y}\right].\nn
\eea
Note however that $m_c, m_d\sim \mathcal{O}(m_{H_i},m_{\eta})$ in (B) which does not apply in the case of Ref.~\cite{Babu:2002uu}.  To estimate this new contribution, we set $t_1=t_2=t$, and obtain
\bea
\widetilde{I}_{acbd}\simeq\dfrac{1.52}{(16\pi^2)^2m_{\eta}^2r_i}.\nn
\eea
The neutrino mass in (B) is then 
\bea
(M_{\nu})_{ab}=M_{ab}+ M^\prime_{ab}.\nn
\eea
Our numerical result shows that the two terms $M_{ab}$ and $M^\prime_{ab}$ are of the same order, and the neutrino mass matrix is sensitive to the mass splitting between $H_2$ and $H_1$. For illustration we choose $r_1=1$, $r_2=0.8$, $\sin\alpha=0.1$, $f_{ac}=f_{bd}=h_{cd}=\lambda=0.05$. That yields $M_{\nu}\sim 0.1~{\rm eV}$.

\bibliographystyle{apsrev}
\bibliography{reference}

\end{document}